\documentclass[epj]{webofc}
\usepackage[utf8]{inputenc}
\usepackage[varg]{txfonts}   
\usepackage{booktabs}
\usepackage{xcolor}
\definecolor{darkred}{rgb}{0.4,0.0,0.0}
\definecolor{darkgreen}{rgb}{0.0,0.4,0.0}
\definecolor{darkblue}{rgb}{0.0,0.0,0.4}
\usepackage[bookmarks,linktocpage,colorlinks,
    linkcolor = darkred,
    urlcolor  = darkblue,
    citecolor = darkgreen]{hyperref}
\usepackage{subfigure}
\wocname{EPJ Web of Conferences}
\woctitle{Lattice2017}
\begin{document}
%
\selectlanguage{english}
\title{Flavor-singlet spectrum in multi-flavor QCD}
\author{%
\firstname{Yasumichi} \lastname{Aoki} \inst{1,2} \and
\firstname{Tatsumi} \lastname{Aoyama} \inst{3} \and
\firstname{Ed}  \lastname{Bennett} \inst{4} \and
\firstname{Masafumi} \lastname{Kurachi} \inst{1,5} \and
\firstname{Toshihide} \lastname{Maskawa} \inst{6} \and
\firstname{Kohtaroh} \lastname{Miura} \inst{7} \and
\firstname{Kei-ichi} \lastname{Nagai} \inst{6} \and
\firstname{Hiroshi} \lastname{Ohki} \inst{8} \and
\firstname{Enrico} \lastname{Rinaldi} \inst{2,12}\fnsep\thanks{Speaker, \email{erinaldi@bnl.gov}\\
Slides of the presentation are available at \url{https://makondo.ugr.es/event/0/session/96/contribution/350}\\
{\bf Preprint numbers:} KEK-CP-362, RBRC-1255.}
\and
\firstname{Akihiro} \lastname{Shibata} \inst{9}  \and
\firstname{Koichi} \lastname{Yamawaki} \inst{6} \and
\firstname{Takeshi} \lastname{Yamazaki} \inst{10,11}
\firstname{    (LatKMI Collaboration)}
}
\institute{%
Institute of Particle and Nuclear Studies, High Energy Accelerator Research Organization (KEK), Tsukuba 305-0801, Japan
\and
RIKEN BNL Research Center, Brookhaven National Laboratory, Upton, NY, 11973, USA
\and
Yukawa Institute for Theoretical Physics, Kyoto University, Kyoto 606-8502, Japan
\and
College of Science, Swansea University, Singleton Park, Swansea, SA2 8PP, UK
\and
Research and Education Center for Natural Sciences, Keio University,
Hiyoshi 4-1-1, Yokohama, Kanagawa 223-8521, Japan
\and
Kobayashi-Maskawa Institute for the Origin of Particles and the Universe, Nagoya University, Nagoya 464-8602, Japan
\and
Centre de Physique Theorique(CPT), Aix-Marseille University, Campus de Luminy, Case 907, 163 Avenue de Luminy, 13288 Marseille cedex 9, France
\and
Department of Physics, Nara Women’s University, Nara 630-8506, Japan
\and
Computing Research Center, High Energy Accelerator Research Organization (KEK), Tsukuba 305-0801, Japan
\and
Faculty of Pure and Applied Sciences, University of Tsukuba, Tsukuba, Ibaraki 305-8571, Japan
\and
Center for Computational Sciences, University of Tsukuba, Tsukuba, Ibaraki 305-8577, Japan
\and
Nuclear Science Division, Lawrence Berkeley National Laboratory, Berkeley, CA, 94720, USA
}
\abstract{%
Studying SU(3) gauge theories with increasing number of light fermions is relevant both for understanding the strong dynamics of QCD and for constructing strongly interacting extensions of the Standard Model (e.g. UV completions of composite Higgs models).
In order to contrast these many-flavors strongly interacting theories with QCD, we study the flavor-singlet spectrum as an interesting probe.
In fact, some composite Higgs models require the Higgs boson to be the lightest flavor-singlet scalar in the spectrum of a strongly interacting new sector with a well defined hierarchy with the rest of the states.
Moreover, introducing many light flavors at fixed number of colors can influence the dynamics of the lightest flavor-singlet pseudoscalar.
We present the on-going study of these flavor-singlet channels using multiple interpolating operators on high-statistics ensembles generated by the LatKMI collaboration and we compare results with available data obtained by the Lattice Strong Dynamics collaboration.
For the theory with 8 flavors, the two collaborations have generated configurations that complement each others with the aim to tackle the massless limit using the largest possible volumes.
}
\maketitle
\section{Introduction}\label{intro}


We perform lattice field theory simulations of QCD with different numbers of light (massless) fermions to study the spectrum of bound states in the scalar and pseudoscalar channel, in particular focusing on particles that have no flavor charge (flavor singlets).
The flavor-singlet scalar channel has the same symmetries of the QCD vacuum and it is related to the trace anomaly, while the flavor-singlet pseudoscalar channel is related to the chiral anomaly.
These anomalies dictate the low-energy behavior of QCD and are instrumentals in the construction of effective models, hence a direct study of these flavor-singlet states is of paramount importance for the understanding of strong dynamics.

While experiments provide direct access to this spectrum in the case of QCD with $N_f=2$--where only the up and down flavors are light-- we can only use approximate models with somewhat large theoretical uncertainties to understand the spectrum for $N_f>2$.
Our lattice simulations aim at filling this gap and at providing data with which phenomenological models could be benchmarked.

This is not only relevant for the understanding of strongly interacting theories like QCD, but also for physics Beyond the Standard Model (BSM)---in particular extensions of the Standard Model based on strong dynamics.
For example, composite Higgs models regard the SM Higgs sector as a low-energy description of new strong dynamics where the Higgs boson is a flavor-singlet scalar bound state.
Therefore, the study of this channel with first-principles calculations becomes important to determine which UV completion has the correct properties to behave like the SM Higgs sector at low energies.

Composite Higgs models based on the idea of \emph{walking technicolor} require that the strongly interacting theory has near-conformal dynamics and these have been studied on the lattice in recent years~\cite{svetitskyPlenary,Pica:2017gcb}.
Interestingly, it is still not clear what would the signs of near-conformal dynamics be at the full non-perturbative level even though there are qualitative expectations from approaches like SD ladder equations which capture some features of the non-perturbative dynamics.
These are only supposed to be used as guidelines for the lattice investigation.
In order to shed light on signals of near-conformal dynamics we compare the flavor-singlet spectrum from lattice simulations of QCD with $N_f=4$, 8 and 12.

Previous lattice numerical simulations indicate that these 3 different values of $N_f$ correspond to rather different type of dynamics: while $N_f=4$ is expected to show the same features of QCD with $N_f=2$, namely confinement and spontaneous chiral symmetry breaking, the theory with $N_f=12$ is expected to have conformal dynamics, with ratios of mass scales that stay constant toward the massless limit.
This points to QCD with $N_f=8$ as a candidate for near-conformal dynamics, with several lattice studies confirming the presence of a flavor-singlet scalar state much lighter than the vector state, in contrast with QCD with smaller number of flavors\footnote{So far we have used the term QCD to refer to a generic SU(3) gauge theory with $N_f$ light quarks in the fundamental irreducible representation of the gauge group, and we continue to do so in the following.}.

In the following we will highlight the differences, or the similarities, between QCD theories with different numbers of light (or massless) flavors, in a quantitative way.
We leave the interpretation of the results in terms of composite Higgs models, or walking technicolor theories, to future and more thorough investigations\footnote{We also remind the readers that some of the unpublished results reported in the following sections should be considered as preliminary and might change (albeit slightly) in the future.}.

\section{Setup of lattice simulations}\label{sec:lattice-setup}

\begin{figure}[thb]
  \centering
  \sidecaption
  \includegraphics[width=0.475\textwidth,clip]{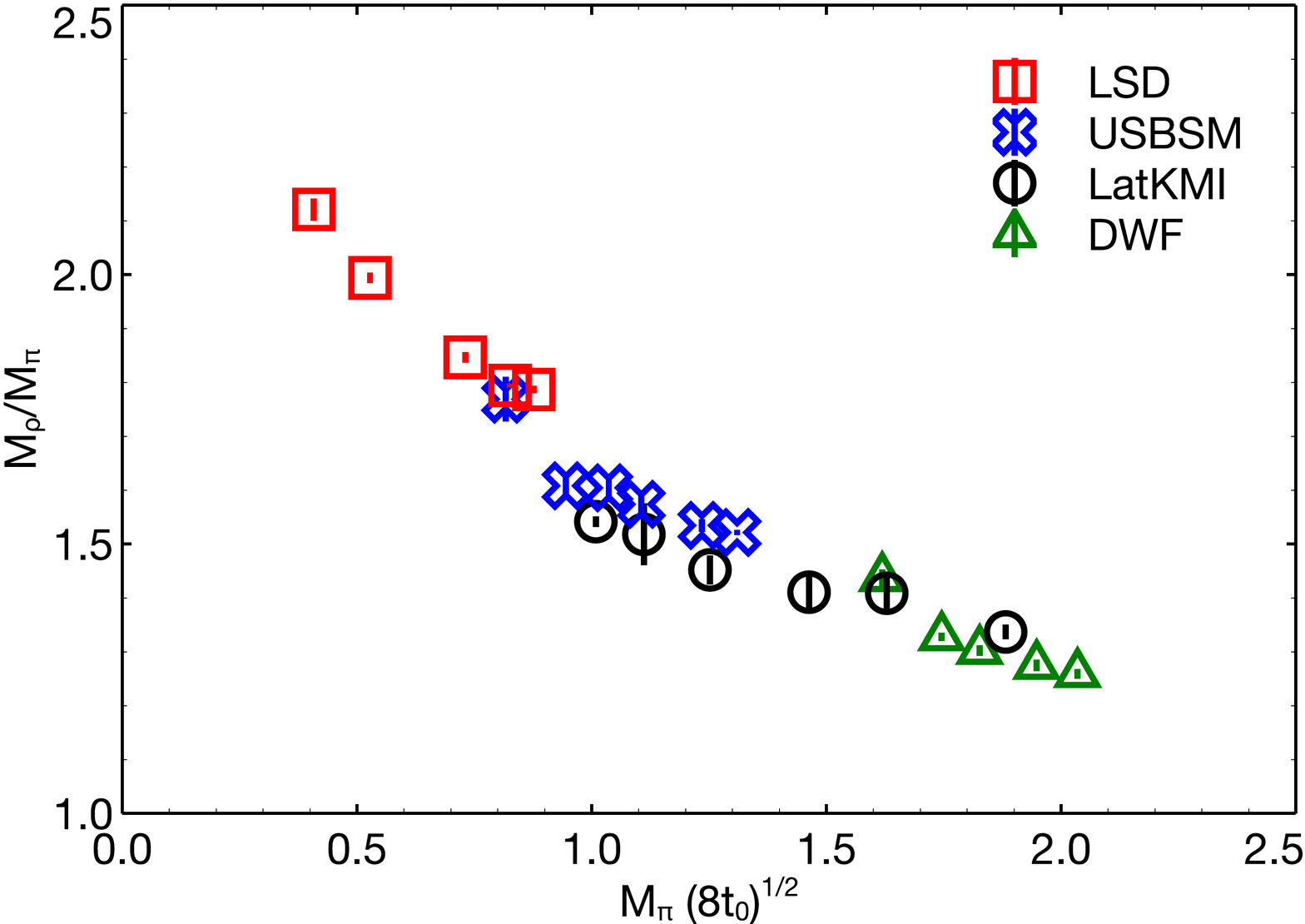}
  \caption{The ratio $M_{\rho}/M_{\pi}$  shows indications of spontaneous chiral symmetry breaking in the massless fermion limit for QCD with $N_f=8$. The results from several lattice studies of different collaborations are shown as a function of the pion mass in units of the Wilson flow scale $\sqrt{8t_0}$, evaluated at each point. See text for references. Image from Ref.~\cite{Appelquist:2016viq}.}
  \label{fig:chisb}
\end{figure}

The LatKMI collaboration has been systematically investigating QCD with many fundamental fermions $N_f=4$, 8, 12, and 16 using a common setup for the lattice action, at several values of the gauge coupling, fermion mass and volume size~\cite{Aoki:2012eq,Aoki:2013xza,Aoki:2013zsa,Aoki:2014oha,Aoki:2016wnc}.
The fermionic action discretization used is the highly improved staggered quark (HISQ) action, while the gauge action is tree-level improved Symanzik.
This setup is advantageous when trying to reach the chiral limit, because it suppresses the effects of taste breaking which spoil flavor symmetry, compared to other commonly used staggered quark actions~\cite{Aoki:2013xza,Aoki:2016wnc}.

In the following, we will compare some of our results for QCD with $N_f=8$, to the findings of the Lattice Strong Dynamics (LSD) collaboration.
They have been using a different lattice discretization to study the spectrum of the 8-flavor theory which is based on the asqtad staggered action with a mixed-representation plaquette action.
With such a discretization, they have been able to access much lighter fermion masses on large volumes, because they can use a somewhat smaller gauge coupling corresponding to a larger lattice spacing~\cite{Appelquist:2016viq}.



The spectrum of the low-lying flavor-non-singlet states is extracted using standard techniques involving the analysis of two-point correlation functions of interpolating operators with the correct quantum numbers.
The benchmark states are the vector and the pseudoscalar ones, which are called $\rho$ meson and the pion $\pi$.
Their masses are easy to compute and their mass ratio is a good indicator for spontaneous chiral symmetry breaking or for conformality.
In the first case, the $\pi$ mass would be going to zero and $M_{\rho}/M_{\pi}$ would go to infinity in the massless fermion limit.
On the other hand, an infrared conformal spectrum in the presence of an explicit fermion mass would show a constant $M_{\rho}/M_{\pi}$ towards the masslesss limit, with possible corrections to this behavior at large quark masses.
For QCD with $N_f=4$ and 8, the lattice simulations show a decisive increase of the ratio $M_{\rho}/M_{\pi}$ towards the massless quark limit.
The interesting case of $N_f=8$ is shown in Fig.~\ref{fig:chisb}, where results from different collaborations are included.
In the plot we see how DWF simulations~\cite{Appelquist:2014zsa} only explore the larger pion mass region, followed by LatKMI results with HISQ fermions~\cite{Aoki:2016wnc}, while results with staggered fermions from the USBSM~\cite{Schaich:2013eba} and LSD collaborations~\cite{Appelquist:2016viq} reach the lightest fermion masses, where $M_{\rho}/M_{\pi}>2$.

\section{The flavor-singlet scalar state}\label{sec:scalar}

\begin{figure}[tp]
   \centering
  \subfigure{\includegraphics[width=0.45\textwidth,clip]{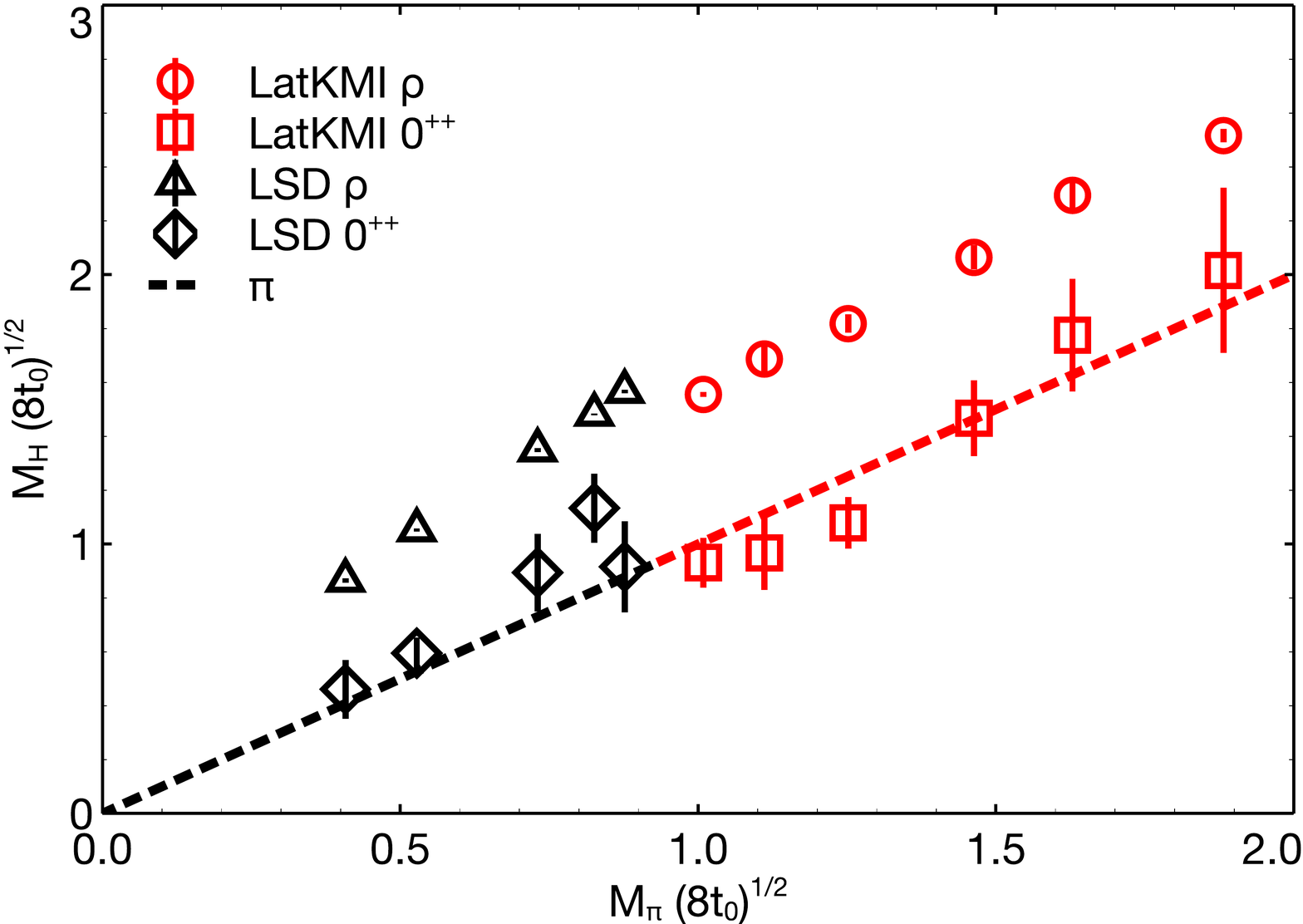}}\hfill
   %
  \subfigure{\includegraphics[width=0.45\textwidth,clip]{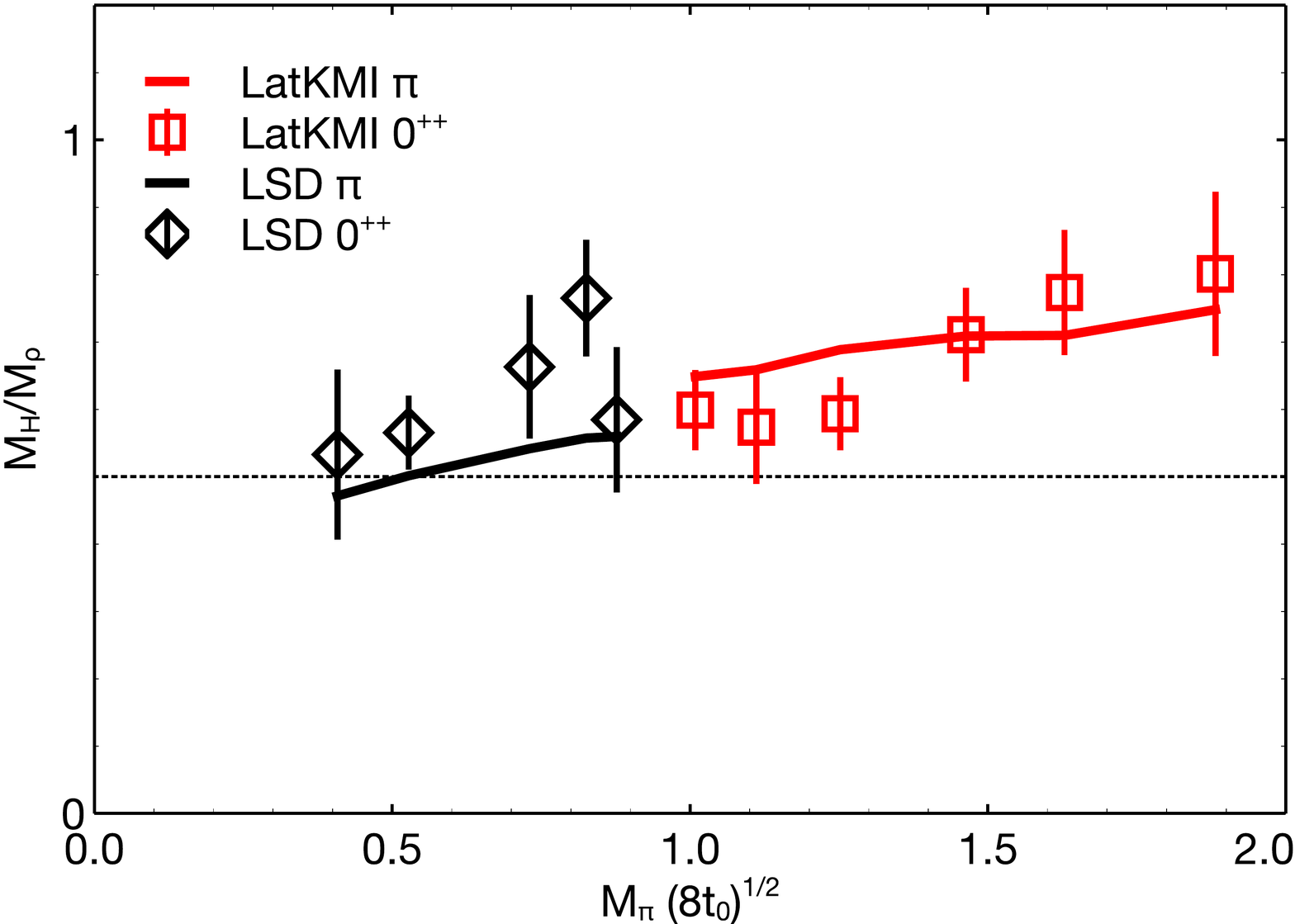}}
   \caption{Combining LatKMI and LSD results for the flavor-singlet scalar state, a clear trend can be seen towards the masslesss fermion limit. The $0^{++}$ state remains degenerate with the pion (left) and much lighter than the vector meson (right), all the way to the lightest fermions studied by the LSD collaboration~\cite{flemingTalk}.}
   \label{fig:scalar-all}
\end{figure}

Extracting the mass of the flavor-singlet scalar state with $J^{PC}=0^{++}$ is difficult because of the need to calculate expensive disconnected contributions and to tame the statistical noise from gauge fluctuations.
Even from the experimental point of view it is hard to study the lightest $0^{++}$ properties because this state appears only as a broad resonance, identified as the $f_0(500)$ or $\sigma$ particle with mass $\sim 400-500$ MeV~\cite{Patrignani:2016xqp}.
In that case the $\sigma$ decays to two pions.
This is an additional complication for lattice calculations---even for the real-world case of QCD with $N_f=2$ light fermions, results have appeared only in the last year~\cite{Briceno:2016mjc}.
The lattice results support the nature of the $\sigma$ as a broad resonance if the quark mass is not too heavy.
However, if the pion in QCD with $N_f=2$ becomes heavier, the $\sigma$ does not decay and it becomes a bound state with a mass $M_{\sigma} \approx 0.9M_{\rho}$ when $M_{\rho}/M_{\pi}\approx 2.2$~\cite{Briceno:2016mjc}, or even heavier than the $\rho$ meson, at larger fermion masses~\cite{Kunihiro:2003yj}.

In this section we would like to highlight the difference between the $N_f=2$ theory and the large $N_f$ theories.
For example, our most precise and extensive results in QCD with $N_f=8$ show that the lightest $0^{++}$ particle is degenerate with the pion in the whole fermion mass range explored.
This of course simplifies the numerical analysis with respect to the real-world scenario because the states propagating in the two-point function are stable.
Moreover, not only the lightest $0^{++}$ state is degenerate with $\pi$, but it is also much lighter than the $\rho$ meson~\cite{Aoki:2016wnc}.

The LSD collaboration has performed the same measurements, with a more sophisticated analysis technique, for QCD with $N_f=8$ using a different discretization, as described in Sec.~\ref{sec:lattice-setup}.
While the hadron masses in units of the lattice spacing are different, using the Wilson flow scale $\sqrt{8t_0}$~\cite{Luscher:2010iy} measured on every ensemble\footnote{For LatKMI ensembles, the Wilson flow scale is reported in Ref.~\cite{Aoki:2016wnc}, while for LSD ensembles we use results that have been shared in private communications and used in Ref.~\cite{Appelquist:2016viq}.} to rescale all results allows us to combine the two datasets and draw suggestive comments.
In Fig.~\ref{fig:scalar-all} the LatKMI results~\cite{Aoki:2016wnc} and the LSD results~\cite{Appelquist:2016viq,flemingTalk} are combined.
The left panel shows the flavor-singlet scalar state following the pion from heavier (LatKMI) to lighter (LSD) fermion masses.
In the fermion mass region explored there is no sign of the $0^{++}$ state ``peeling'' off from the pion, which would happen at very light fermion masses due to the pseudo-Nambu-Goldstone nature of the pions.
The right panel instead shows a direct comparison between the flavor-singlet scalar state and the $\rho$ meson.
There is a clear trend towards the massless fermion limit, indicating that the mass hierarchy between the two states increases.
As a comparison, the lattice results for QCD with $N_f=2$ for a pion to vector mass ratio corresponding to the lightest point in the figure yield $M_{\sigma} \approx 0.9M_{\rho}$~\cite{Briceno:2016mjc}.

We can contrast the $N_f=8$ flavor-singlet scalar result with the $N_f=12$ case, which was the first discovered example of a many-flavor theory showing a light scalar state~\cite{Aoki:2013zsa}.
For QCD with $N_f=12$ we have found a flavor-singlet state even lighter than the pion.
Moreover, its mass interestingly follows a hyperscaling relation dictated by the mass anomalous dimension $\gamma$ (extracted from scaling relations for the rest of the spectrum, which has much smaller statistical uncertainties).

We also mention that we have extracted the mass of the flavor-singlet scalar state in QCD with $N_f=4$ and it's mass on our lightest point is heavier than the pion mass but lighter than the vector mass~\cite{Aoki:2015zny}.
This result is shown in panel (a) of Fig.~\ref{fig:spectrum-all}.

The qualitative conclusion we draw from the wealth of our results is that the flavor-singlet scalar state gets lighter with respect to the $\rho$ and $\pi$ states as the number of light degenerate flavors increases (eventually becoming the lightest state in the spectrum when the theory enters the conformal window).

\section{The flavor-singlet pseudoscalar state}\label{sec:pseudoscalar}

While the results presented for the flavor-singlet scalar had been previously presented in conference talk or in published papers, the flavor-singlet pseudoscalar results were presented for the first time at this conference for three values of $N_f=4$, 8 and 12.
Preliminary results for $N_f=8$ have been reported in Ref.~\cite{Aoki:2016fxd}.

The flavor-singlet pseudoscalar state has quantum numbers $J^{PC}=0^{-+}$ and corresponds to the so-called $\eta^\prime$ particle in QCD with $N_f=2$.
Witten~\cite{Witten:1979vv} and Veneziano~\cite{Veneziano:1980xs} showed that the $\eta^\prime$ mass is directly related to the contribution of the axial anomaly and the topological structure of the QCD vacuum.
In fact, in the limit of a SU($N_c$) gauge theory with $N_c\rightarrow \infty$, the $\eta^\prime$ becomes massless, behaving as a Nambu-Goldstone boson.
This mechanism for $N_c=3$, therefore for QCD, is not applicable and the axial anomaly makes this state very heavy, measured experimentally to be $\sim 958$ MeV~\cite{Patrignani:2016xqp}.

It is notoriously difficult to compute the mass of the flavor-singlet pseudoscalar, because the pion contribution is statistically challenging to remove.
In our calculations we adopt a gluonic operator, which does not suffer from the aforementioned problem, since it does not couple directly to $\pi$ states.
The same method has been adopted in QCD with $N_f=2+1$ and has led to results in agreement with experiments~\cite{Fukaya:2015ara}.

In practice, the interpolating operator used is the topological charge density defined through the clover-plaquette strength-energy field tensor  $F^{\mu \nu}(x)$:
\begin{equation}
  \label{eq:qtopo}
  q(x) \ = \ \frac{1}{32\pi} \epsilon_{\mu \nu \rho \sigma} \textrm{Tr}\, F^{\mu \nu}(x) F^{\rho \sigma}(x) \quad ,
\end{equation}
The two-point function $\langle q(x) q(y) \rangle$ is computed for all pairs of points ($x$,$y$) in the four-dimensional volume $L^3 \times T$ efficiently using FFT.
Moreover, because of translation invariance, the two-point correlator only depends on the distance $r=|x-y|$ and we average all contributions at fixed distance to increase statistics.
For a particle freely propagating in four dimensions, the correlator takes the form
\begin{equation}
  \label{eq:correlator2}
  C(r)=\frac{A}{r^{1.5}}\left(1+\frac{3}{8r}\right)e^{-M_{\eta^\prime}r} 
\end{equation}
at large distances $r \rightarrow \infty$.
We fit the data of $C(r)$ to this form, in a specific window of distances $r \in [r_{\rm min},r_{\rm max}]$, to extract the two parameters $A$ and $M_{\eta^\prime}$.

Additionally, we utilize the Wilson flow~\cite{Luscher:2010iy} as a smearing technique to remove ultraviolet fluctuations and obtain interpolating operators with \emph{physical size} --- having enhanced overlap to the ground state.
Using the operator in Eq.~(\ref{eq:qtopo}), where $F^{\mu \nu}(x)$ is computed for several Wilson flow times $t_w$, we obtain a large number of correlators $C_{t_w}(r) = -\langle q_{t_w}(x) q_{t_w}(y) \rangle$.
The statistical fluctuations are dramatically reduced by the Wilson flow smearing, such that correlators at larger $t_w$ can be easily fitted to the exponential form in Eq.~(\ref{eq:correlator2}).
However, the smearing introduces systematic corrections that have to be addressed~\cite{Bruno:2014ova}.
In fact, it turns out that the dominant source of uncertainty in extracting $M_{\eta^\prime}$ is coming from systematic effects of the fitting procedure.
There are two competing effects:
\begin{itemize}
\item Eq.~(\ref{eq:correlator2}) can only be assumed to be valid in a specific region of large $r$, where only the ground state dominates, such that the extracted mass does not depend on the value of $r_{\rm min}$.
\item The correlator at large distances suffers from larger statistical fluctuations and can be extracted only at large values of smearing $t_w$, where smearing artifacts~\cite{Bruno:2014ova} are larger.
\end{itemize}

\begin{figure}[thb]
  \centering
  \sidecaption
  \includegraphics[width=0.575\textwidth,clip]{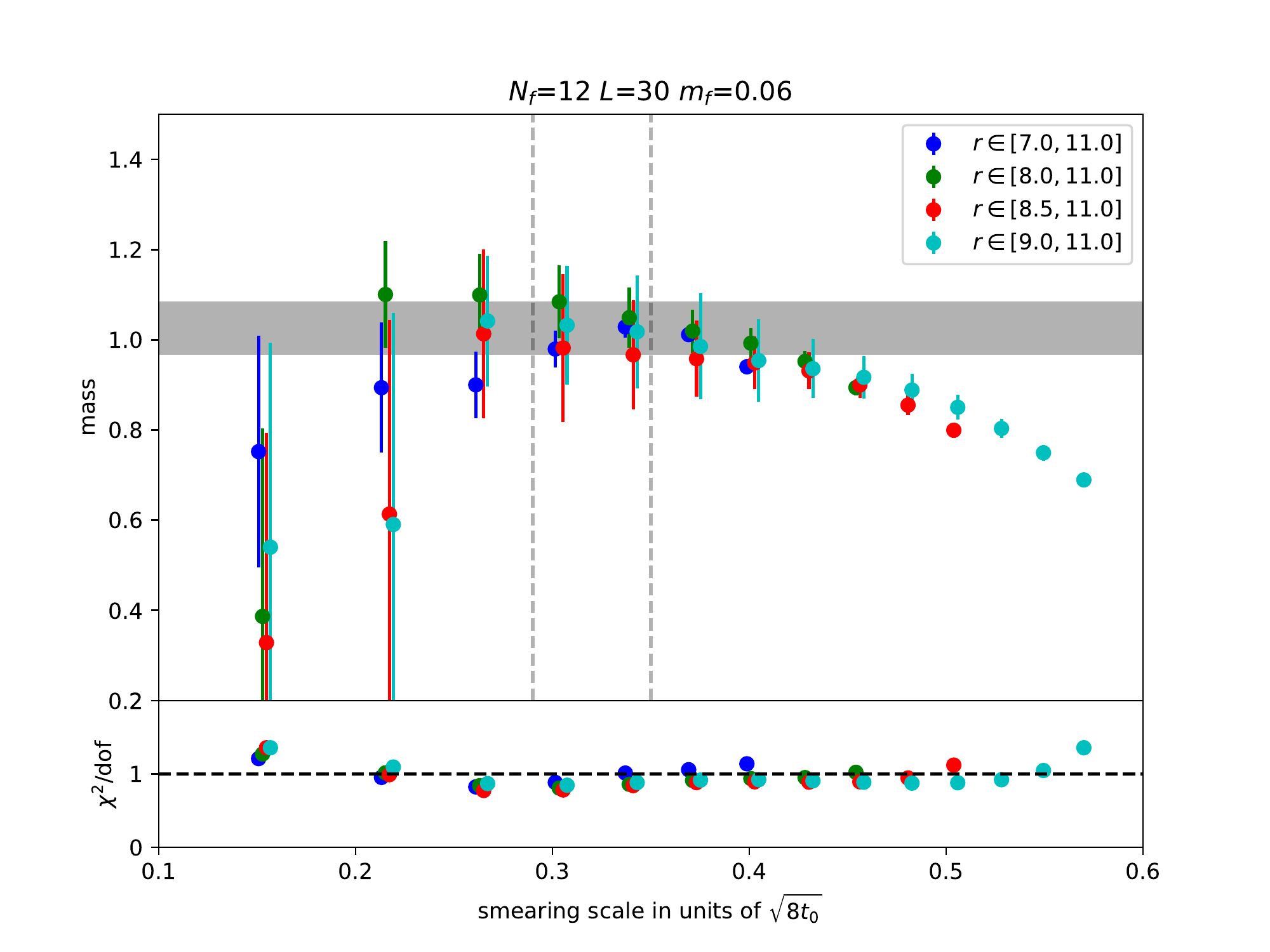}
  \caption{The $\eta^\prime$ mass fitted for different distance regions and smearings, for a specific ensemble of QCD with $N_f=12$. The main features are described in the text.}
  \label{fig:eta-prime-syst}
\end{figure}

We estimate the systematic uncertanity on $M_{\eta^\prime}$ by looking for a plateaux in the fitting range $[r_{\rm min},r_{\rm max}]$, and for a plateaux in the smearing range $\sqrt{8t_w}$.
This is exemplified in the panels of Fig.~\ref{fig:eta-prime-syst} for a representative ensemble of QCD with $N_f=12$.
The smearing range $\sqrt{8t_w}$ is considered in units of the characteristic radius given by the Wilson flow scale $\sqrt{8t_0}$, and we find a common region for all the ensembles at fixed $N_f$---we see that it does not depend on the fermion mass or the volume.
This indicates that such a smearing corresponds to some physical scale for the operator with the best coupling to the ground state.
In the identified region, we take the difference between the largest and the smallest fitted mass as an estimate of the systematic error.
The statistical errors of the individual points are usually smaller than this systematic uncertainty.
In Tab.~\ref{tab:eta-errors} we report the $\eta^\prime$ masses with their errorbars estimated as reported above, on all our ensembles.

\begin{table}[thb]
  \small
  \centering
  \caption{Results for the $\eta^\prime$ mass in QCD with $N_f=$4, 8 and 12. The error is dominated by systematic uncertainties of the fitting procedure on the correlator described in Sec.~\ref{sec:pseudoscalar}.}
  \label{tab:eta-errors}
  \begin{tabular}{lll|lll|lll}\toprule
    \multicolumn{3}{c}{$N_f=4$} & \multicolumn{3}{|c|}{$N_f=8$} & \multicolumn{3}{c}{$N_f=12$} \\ \midrule
   $L$ & $m_f$ & $\eta^\prime$ & $L$ & $m_f$ & $\eta^\prime$ & $L$ & $m_f$ & $\eta^\prime$ \\
    20 & 0.01 & 0.741(39) & 42 & 0.012 & 0.875(55) & 30 & 0.04 & 0.815(43)\\
    20 & 0.02 & 0.746(37) & 36 & 0.015 & 0.954(63) & 30 & 0.05 & 0.850(44)\\
    20 & 0.03 & 0.788(33) & 36 & 0.02  & 0.956(49) & 30 & 0.06 & 1.025(59)\\
    20 & 0.04 & 0.794(36) & 30 & 0.03  & 0.945(69) &  \multicolumn{3}{c}{} \\
    \multicolumn{3}{c|}{} & 30 & 0.04  & 0.977(42) &  \multicolumn{3}{c}{} \\\bottomrule
  \end{tabular}
\end{table}

\begin{figure}[bht]
   \centering
   \subfigure{\includegraphics[width=0.33\textwidth,clip]{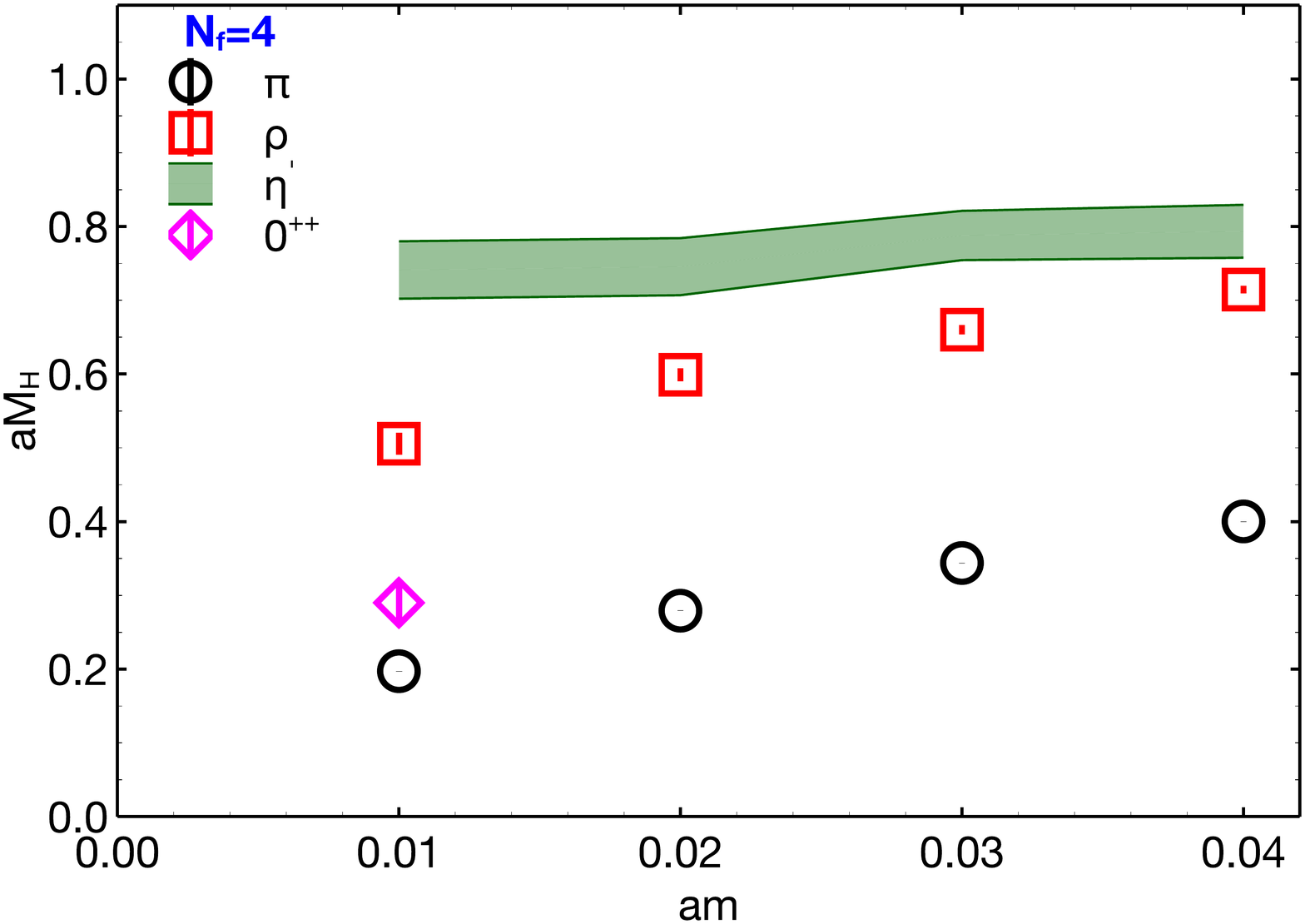}}\hfill
   %
  \subfigure{\includegraphics[width=0.33\textwidth,clip]{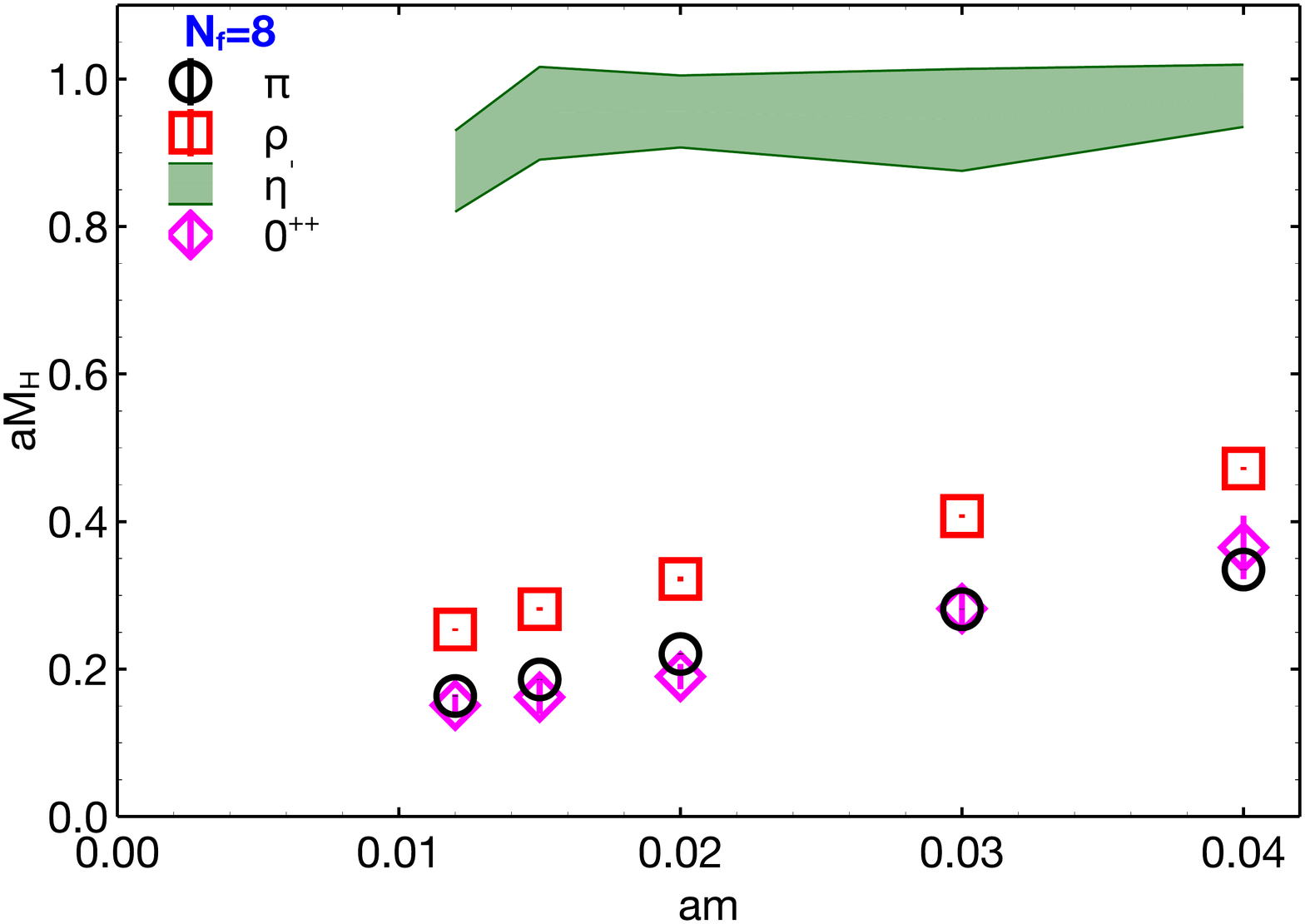}}\hfill
   %
  \subfigure{\includegraphics[width=0.33\textwidth,clip]{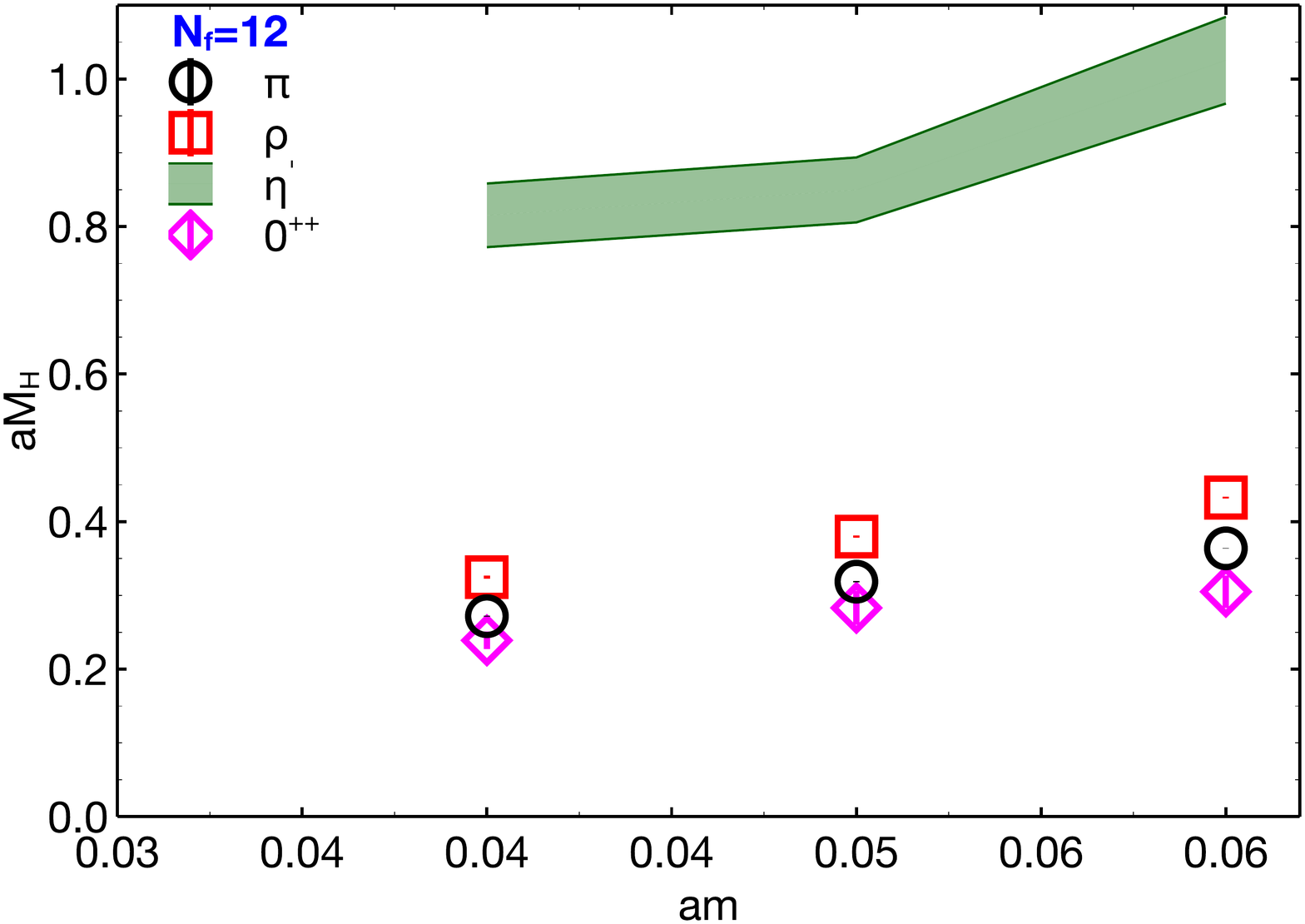}}\hfill
   \caption{Flavor-singlet scalar and pseudoscalar spectrum compared to flavor-non-singlet pseudoscalar and vector spectrum for $N_f=4$, 8 and 12. The $\eta^\prime$ mass is shown with error bands that reflect a large systematic uncertainty. Only statistical errors are shown for the other states.}
   \label{fig:spectrum-all}
\end{figure}

We can now compare the flavor-singlet pseudoscalar state with the rest of the low-lying spectrum as we change the number of flavors.
The compilation of results for QCD with $N_f=4$, 8 and 12 is reported in Fig.~\ref{fig:spectrum-all}.
We identify a notable increase in the gap between the flavor-singlet pseudoscalar and the vector meson.
In QCD with $N_f=4$ the mass ratio $M_{\eta^\prime}/M_{\rho}$ is close to one, while it grows to $\sim 3-4$ for $N_f=8$ and 12.

\begin{figure}[thb]
  \centering
  \sidecaption
  \includegraphics[width=0.475\textwidth,clip]{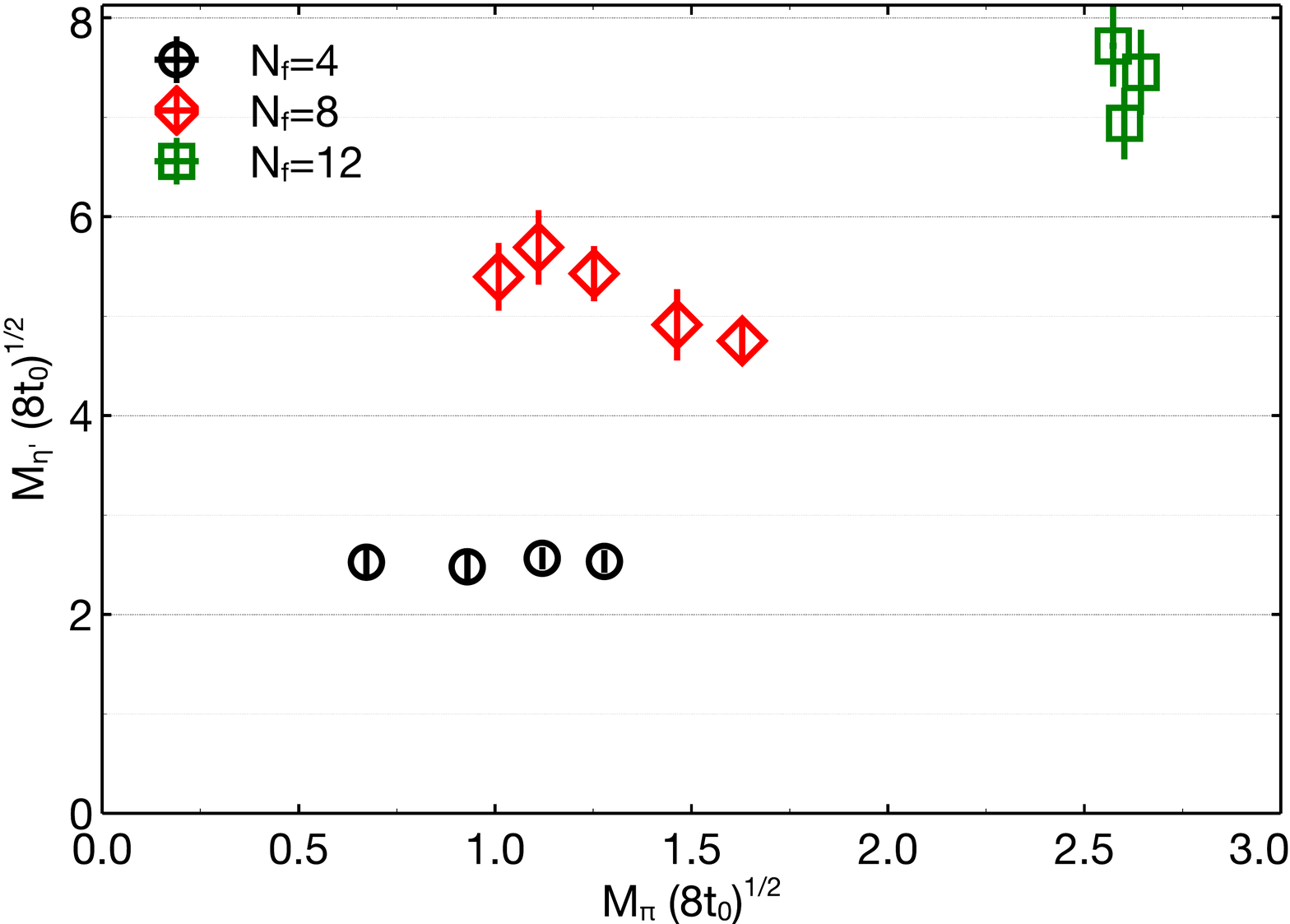}
  \caption{Comparison of the flavor-singlet pseudoscalar mass for $N_f=4$, 8 and 12 as a function of the pion mass. The hadronic masses are in units of the Wilson flow scale $\sqrt{8t_0}$ to partially remove discretization effects due to different gauge couplings for the various $N_f$ theories. Different quark mass regions are explored for different $N_f$ values.}
  \label{fig:eta-prime-all}
\end{figure}

\section{Summary}\label{sec:summary}

There are theoretical expectations~\cite{Matsuzaki:2015sya} suggesting that the flavor-singlet scalar state becomes light near the conformal window as $N_f$ is increased, while the flavor-singlet pseudoscalar becomes heavier.
In particular, the flavor-singlet pseudoscalar is expected to scale with $N_f$ in QCD, as a consequence of a \emph{anti}-Veneziano limit where $N_f/N_c \gg 1$ is fixed as $N_c\rightarrow \infty$.
This is confirmed by comparing QCD with $N_f=4$, 8 and 12 in Fig.~\ref{fig:eta-prime-all} after rescaling all the masses with the Wilson flow scale $\sqrt{8t_0}$.

After the conference, a low-energy description of many-flavor QCD~\cite{Meurice:2017zng}, based on the linear sigma model, was able to incorporate our numerical results for the flavor-singlet scalar and pseudoscalar states in a consistent framework.
This effective description points out a potentially simple way to use our data in order to discriminate if a particular many-flavor QCD theory is inside or outside the conformal window.
Another effective discription based on the linear sigma model was presented at this conference~\cite{gasbarroTalk} and it was benchmarked against the spectrum of QCD with $N_f=8$, along the lines of the works in Refs.~\cite{Appelquist:2017wcg,Appelquist:2017vyy} which used a different effective description including the effects of a dilaton.

In conclusion, understanding the flavor-singlet spectrum of QCD for varying number of flavors from first principles is challenging but it is of paramount importance to obtain a quantitative low-energy description of strong dynamics in different regimes, from spontaneous chiral symmetry breaking to conformality.
\section*{Acknowledgements}
We thank the Lattice Strong Dynamics (LSD) collaboration for sharing their plots and their preliminary unpublished results for the flavor-singlet scalar mass on their ensembles. Numerical calculations have been carried out on the high-performance computing systems at KMI (${\Large\varphi}$), at the Information Technology Center in Nagoya University (CX400), and at the Research Institute for Information Technology in Kyushu University (CX400 and HA8000) through the HPCI System Research Projects (Project ID: hp140152, hp150157, hp160153) as well as the general use.\\
This work is supported by the JSPS Grants-in-Aid for Scientific Research (S) No. 22224003, (C) No. 16K05320 (Y.A.) for Young Scientists (A) No.16H06002 (T.Y.), (B) No.25800138 (T.Y.), (B) No.25800139 (H.O.), (B) No.15K17644 (K.M.), and also by the MEXT Grants-in-Aid for Scientific Research on Innovative Areas No.25105011 (M.K.).
K.M. is supported by the OCEVU Labex (ANR-11-LABX-0060) and the A*MIDEX project (ANR-11-IDEX-0001-02),funded by the ``Investissements d'Avenir'' French government program and managed by the ANR.\\
E.~R. is supported by a RIKEN Special Postdoctoral fellowship.
This work is supported in part by the JSPS KAKENHI Grant No.~16K05320.
\bibliography{lattice2017}

\end{document}